\documentclass[onecolumn,amsmath,amssymb,prd,reprint,longbibliography,floatfix,8pt]{revtex4-1}

\usepackage{natbib}
\usepackage{array}

\usepackage{tabularx} 
\usepackage{float}

\usepackage{mathtools}
\usepackage{cancel}
\usepackage{enumitem}
\usepackage[utf8x]{inputenc}
\usepackage{graphicx}
\usepackage{dcolumn}
\usepackage{bm}
\usepackage[switch]{lineno}
\usepackage{float} 
\usepackage{subfigure}
\usepackage{tikz}
\usepackage{multirow}
\usetikzlibrary{arrows.meta}
\usetikzlibrary{arrows,decorations.pathmorphing,backgrounds,positioning,fit,petri}

\begin{document}

\title{A multireference picture of electronic excited states in vanadyl and copper tetraphenyl porphyrin molecular qubits}
\author{Arup Sarkar,* Alessandro Lunghi}
\email{lunghia@tcd.ie, arsarkar@tcd.ie}

\affiliation{School of Physics, AMBER and CRANN Institute, Trinity College, Dublin 2, Ireland}

\begin{abstract}
The nature of electronic excited states has a deep impact on the dynamics of molecular spins, but remains poorly understood and characterized. Here we carry out a thorough multiconfigurational investigation for two prototypical molecular qubits based on vanadyl and copper tetra-phenyl porphyrins. State-average CASSCF and NEVPT2 calculations have been employed with four different active spaces of growing complexity to account for the d-d, second d-shell, ligand-to-metal charge transfer states and $\pi$-$\pi^*$ excited states, revealing an in-depth picture of low-lying excited states in agreement with experimental observations. The largest active spaces attempted, (13,14) for the vanadyl and (17,12) for the copper compounds, reveal that the lowest-lying excited states originate from $\pi$-$\pi^*$ quartet excitations. These findings shed light on the nature of the excited states of molecular qubits, taking an important step toward elucidating their role in molecular spin dynamics.
\end{abstract}

\maketitle
\section*{Introduction}
\begin{figure}[h!]
  \centering
  \vspace{5pt}
  \includegraphics[width=0.6\textwidth, trim=120 80 100 5, clip]{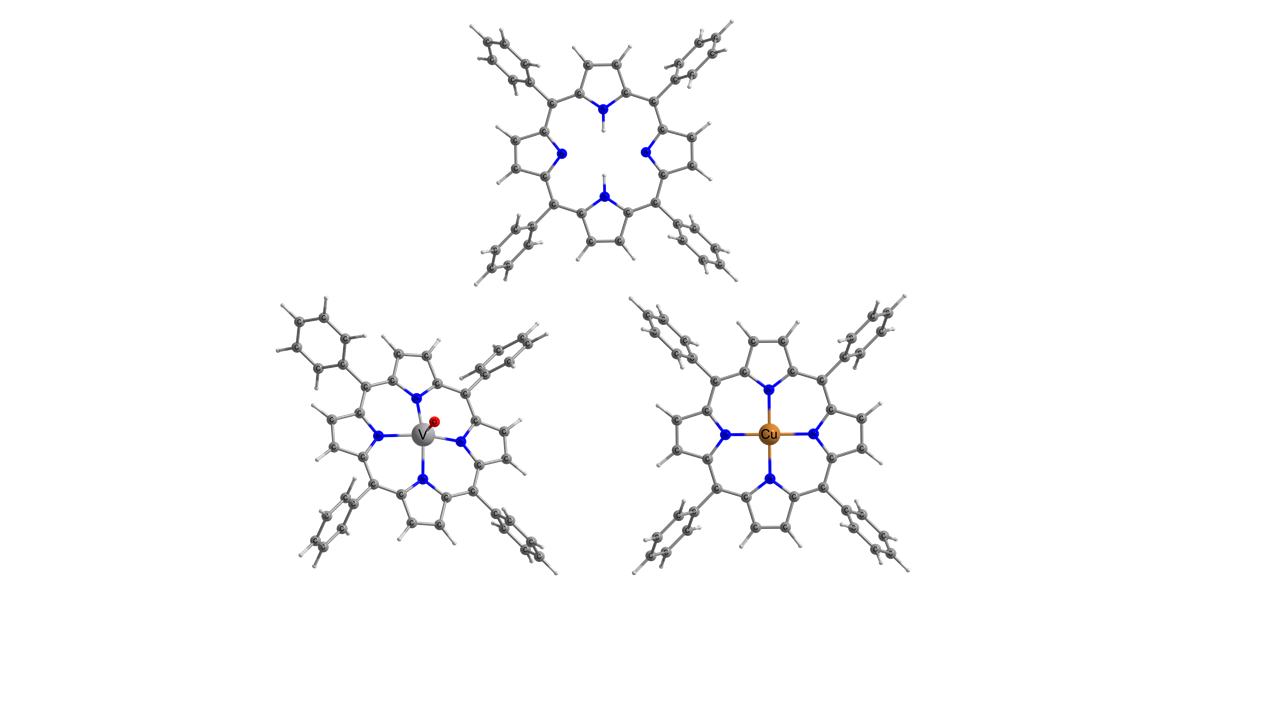}
  \caption{Molecular representation of H$_2$TPP (top), VOTPP (bottom left) and CuTPP (bottom right).}
  \vspace{-10pt}
  \label{fig_mol}
\end{figure}

Porphyrins are planar macrocyclic aromatic compounds with largely delocalized $\pi$-electron clouds widely used in optoelectronics due to their tunable optical and redox properties. Their rigid structure and ability to coordinate with diamagnetic and paramagnetic metal ions alike make them ideal for a range of photophysical applications, including nonlinear optics,\cite{Shi2021CrystallineOptics} light-harvesting,\cite{Mukhopadhyay2018PorphyrinBoxes} chemical/biological sensing,\cite{Bandyopadhyay2025ActivatableTherapeutics} photodynamic therapy\cite{Singh2024ACancer, He2024Anthracene-BasedTherapy} and photocatalytic applications.\cite{Afreen2025Porphyrin-BasedPerspectives} Moreover, porphyrins, and in particular tetraphenylporphyrins (TPPs), exhibit long spin coherence lifetimes when coordinated with spin-1/2 metal ions such as vanadyl (VO$^{2+}$) or Cu$^{2+}$ (see Figure \ref{fig_mol}), making them appealing for quantum information processing applications\cite{Santanni2024MetalloporphyrinsScience, Yu2020SpinQubitsb, Yamabayashi2018ScalingFramework, Eaton2025AnisotropyPorphyrins, Espinosa2025SlowLigand} and building hybrid spin-optical quantum interfaces.\cite{doi:10.1021/jacs.4c10632}

Maximizing the coherence, $T_2$ and lifetime, $T_1$ of molecular spin qubits has so far been a priority and a plethora of spin-1/2 systems containing VO$^{2+}$ or V$^{4+}$ and Cu$^{2+}$ complexes have been investigated over the past decade to interpret the impact of solvents,\cite{Zadrozny2015MillisecondQubit} nuclear spin,\cite{Zadrozny2014MultipleComplex} metal-ligand covalency,\cite{Fataftah2019Metal-ligandCandidates, Follmer2020UnderstandingPhthalocyanines} ligand-field symmetry,\cite{Kazmierczak2021TheCoherence, Kazmierczak2022IlluminatingT1Anisotropy} and d-d excitations\cite{Kazmierczak2025ARelaxationb} on these time constants. Despite the great deal of knowledge produced by these studies, the role of electronic excitations, other than d-d ones, on spin dynamics remains poorly understood. This represents an outstanding knowledge gap with potentially large repercussions on our ability to engineer molecular qubits with improved functionalities. Indeed, both theoretical and experimental evidence for the importance of excited states in determining the relaxation rate of molecule spins has built over the last few years,\cite{Albino2019First-PrinciplesBits,
Kazmierczak2022IlluminatingT1Anisotropy,Kazmierczak2025ARelaxationb} and a recent ab initio study has demonstrated that electronic excited states are key in promoting Raman relaxation starting from already $\sim$20 K and up to room temperature\cite{Mariano2025TheMolecules}. The study of electronic excitations in molecular qubits is made even more urgent considering emerging studies on spin-optical interfaces as a means to initialize and read out molecular spins.\cite{doi:10.1021/jacs.4c10632}

Over the past few decades, the development of high-performance computing has made multiconfiguration (MCSCF) quantum chemical methods a routine choice for describing excited states and correlated wave functions of open-shell systems.\cite{Gagliardi2023SustainableEverywhere, Neese2024APackage} These methods, combined with perturbative corrections (PT2) or other post-CASSCF methods, not only provide an unambiguous description of metal-to-metal electronic excitations, but are also able to accurately estimate magnetic properties such as zero-field splitting, ligand-field interactions, and spin-Hamiltonian parameters in transition metals, lanthanide and actinide complexes.\cite{Atanasov2015FirstMagnets} 
Addressing the true nature of molecular excited states, however, requires going beyond simulating those metal-to-metal transitions. Attempts in incorporating metal-ligand covalency effects have also been done by including a second d-shell\cite{Singh2018ChallengesComplexes} and aromatic/delocalized $\pi$-orbitals \cite{Schlimgen2017StaticDithiolates, Schlimgen2023CharacterizingMethods} in small to medium-sized molecules using various multireference wave function-based approaches. The complexity of such a task is well exemplified by Li Manni et. al.\cite{Manni2016CombiningMetal-Porphyrins} and Zhou et. al.,\cite{Zhou2019MulticonfigurationSpaces}who tested a variety of multiconfiguration methods, such as CASSCF, FCIQMC, and MC-PDFT with different active spaces, to explore the excited states of porphyrin and Fe-porphyrin systems. 

Here, we build on this literature to address the following key aspects of VOTPP and CuTPP's molecular excited states: i) identify the nature of low-lying excited states, ii) determine how the ligand-to-metal charge transfer (LMCT) states, metal-based d-d excitations and ligand-based $\pi$-$\pi^*$ excitations combined describe the absorption spectra, and iii) determine the effect of these excitations on the spin-orbit coupling and $g$-factors. Our work provides an in-depth quantitative depiction of excited electronic states of TPP-based molecular qubits and establishes a solid foundation for spin relaxation theory and the engineering of molecular spin-optical interfaces.

\section*{Computational methods}
All the calculations are performed using the ORCA 5.0.4 quantum chemistry package.\cite{Neese2020ThePackage} Starting geometries of VOTPP and CuTPP are taken from the X-ray structures reported by Yamabayashi et al.,\cite{Yamabayashi2018ScalingFramework} and Urtizberea et al.,\cite{Urtizberea2018ANanosheets} respectively.  Gas-phase geometry optimizations for H$_2$TPP, VOTPP and CuTPP molecules are carried out using BP86\cite{PhysRevB.33.8822} functional with Grimme's D3 dispersion correction\cite{Grimme2010} and Becke-Johnson damping\cite{Grimme2011EffectTheory} function using def2-SVP\cite{Weigend2005BalancedAccuracy} basis set for all the elements. The second-order scalar relativistic Douglas-Kroll-Hess \cite{Wolf2002TheTransformation} Hamiltonian is employed during all the multireference calculations. DKH-contracted basis sets have been used for different atoms- DKH-def2-TZVP\cite {Weigend2005BalancedAccuracy} basis set for the V and Cu centres, DKH-def2-TZVP(-f) for O, N, and C atoms, and DKH-def2-SVP basis set for H. The initial guess orbitals for the CASSCF calculations are generated using restricted open-shell Kohn-Sham (ROKS) or unrestricted KS (quasi-restricted orbitals) calculations. During the SA-CASSCF calculations, different active spaces are chosen for the H$_2$TPP, VOTPP and CuTPP molecules based on the 3d, 4d, and ligand-based $\sigma$ and $\pi$ orbitals (see Table  \ref{tab:Active space composition}). The dimension of the guess matrix was increased to 4000 while dealing with the large active spaces. The N-electron valence state perturbation theory second-order (NEVPT2)\cite{Angeli2001IntroductionTheory} is employed to account for the dynamic electron correlation. Spin-orbit interactions, and subsequent $g$-factors are computed using the quai-degenerate perturbation theory (QDPT) combined with the effective Hamiltonian approach (EHA) using the spin-orbit mean field (SOMF) integral.\cite{Neese2005EfficientCalculations}

 For the H$_2$TPP molecule, the smallest active space (AS1) includes four electrons in five $\pi$ orbitals, i.e., (4,5), comprising HOMO-1, HOMO, LUMO, LUMO+1 and LUMO+2 orbitals. These five $\pi$ orbitals cover the carbons and nitrogen of the central porphyrin ring, excluding the four substituted phenyl groups. In the next step (AS2), two more $\pi$ orbitals are added from the occupied orbitals, making it (8,7) active space (see Table \ref{tab:Active space composition}). The latter two $\pi$ orbitals belongs to the b$_{2g}$ and b$_{3g}$ point group symmetries from the D$_{2h}$ point group which tends to interact with the \textit{d}$_{xz}$ and \textit{d}$_{yz}$ set of the metal 3d orbitals, as discussed later in more details. 

The active spaces for the VOTPP and CuTPP molecules are similarly built using a systematic increase in their complexity. The minimal active space consists of the five metal 3d-orbitals (AS1) and 3d electrons, namely (1,5) and (9,5) for Vanadyl (VO$^{2+}$) and Cu$^{2+}$ ions, respectively. In addition to 3d, the inclusion of the second d-shell, i.e., the 4d (or 3d'), is also tested for these two complexes, leading to the model AS2. Next, in AS3 and AS4, ligand $\sigma$ and $\pi$ orbitals are added to the 3d space to incorporate the ligand-to-metal, metal-to-ligand and ligand-to-ligand excitations. In AS3, a (9,9) active space is chosen for the (VO$^{2+}$) species to accommodate the vanadium 3d and oxygen 2p-based bonding orbitals which constitute d$_{z^2}$-p$_z$ ($\sigma$) and d$_{xz}$-p$_x$/d$_{yz}$-p$_y$ ($\pi$) molecular orbitals (MOs). In the case of CuTPP, this AS3 is restricted to (11,6) by incorporating the bonding counterpart of d$_{{x^2}-{y^2}}$ orbital consisting of the N-2p$_x$/p$_y$ atomic orbitals. The AS4 constitutes the largest active space chosen for these two complexes, which accounts for the delocalized $\pi$-orbitals from the porphyrin ring. In the case of VOTPP, the AS4 or (13,14) consists of all the (9,9) active space orbitals along with the (4,5) active space from TPP MOs. Now, the other two doubly occupied $\pi$ orbitals, which were present in the (8,7) active space of H$_2$TPP, are now not available for bonding with the vanadium d$_{xz}$ or d$_{yz}$ orbitals. For CuTPP, AS4 consists of (9,5) with the (4,5) active space from H$_2$TPP and two more doubly occupied $\pi$ orbitals from the porphyrin ring, making it (17,12). These two active spaces are certainly not the largest possible active space for this system; however, from the standpoint of state-average calculations with low-lying excited states, it can be considered an optimal choice within the CASSCF limit.

\begin{table}[]
\centering
\caption{\textbf{Active space composition of three molecules}}
\label{tab:Active space composition}
\renewcommand{\arraystretch}{1.1}
\resizebox{\columnwidth}{!}{%
\begin{tabular}{ccccc}
\hline
      & AS1   & AS2    & AS3    & AS4     \\ \hline
H$_2$TPP & (4,5) & (8,7)  & -      & -       \\ \hline
 & 3d & 3d + 4d & \begin{tabular}[c]{@{}c@{}}3d + \\ 1st coordination\\ sphere $\sigma$/$\pi$\end{tabular} & \begin{tabular}[c]{@{}c@{}}3d +\\  delocalized $\pi$\end{tabular} \\ \hline
VOTPP & (1,5) & (1,10) & (9,9)  & (13,14) \\ \hline
CuTPP & (9,5) & (9,10) & (11,6) & (17,12) \\ \hline
\end{tabular}%
}
\renewcommand{\arraystretch}{1.0}
\end{table}

\section*{Results and Discussion}

\begin{figure}[h]
  \centering
  \vspace{5pt}
  \includegraphics[width=1.40\columnwidth, trim=120 10 10 20, clip]{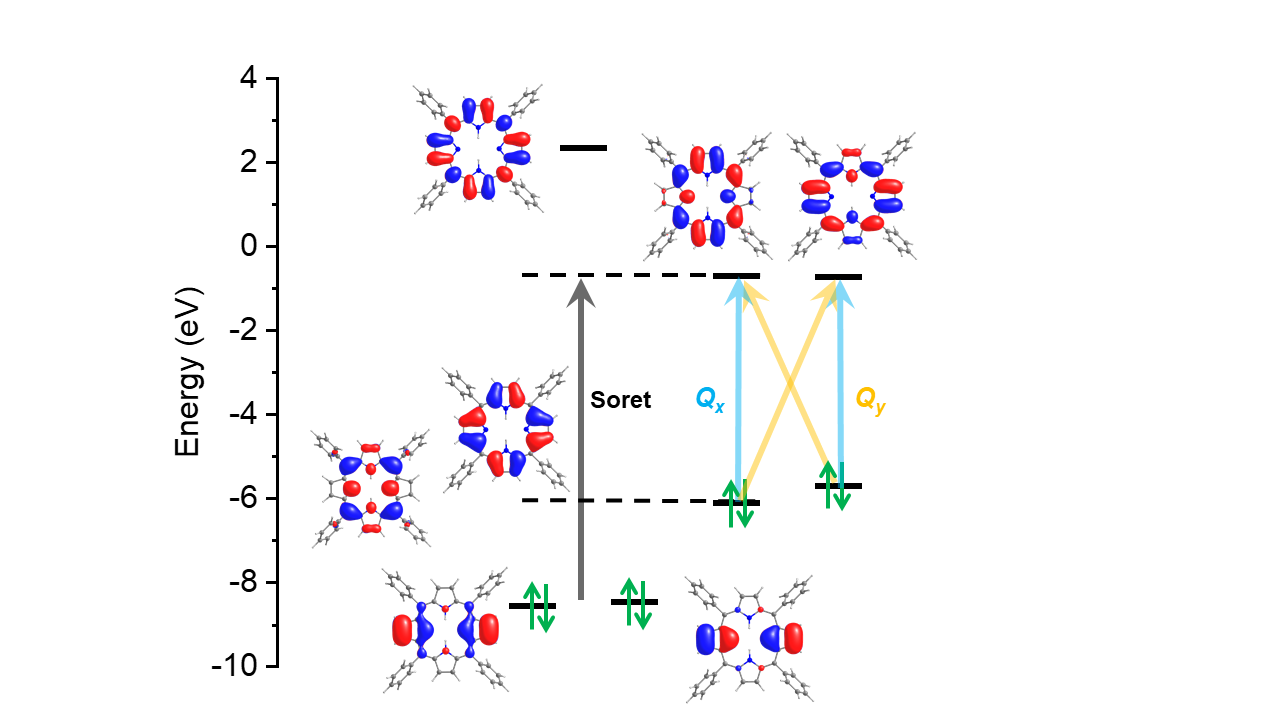}
  \caption{Absolute energies of the state-average orbitals of H$_2$TPP obtained from CASSCF(8,7) calculations. The electronic occupation corresponds to the ground-state electron configuration. The four orbitals are HOMO, HOMO-1 and LUMO, LUMO+1.}
  \vspace{-10pt}
  \label{fig_2}
\end{figure}

\textbf{Free tetra-phenyl porphyrin ring.} At first, the electronic excited states for the neutral ligand framework are examined. The ground state of the free tetra-phenyl porphyrin (H$_2$TPP) is a closed-shell singlet with valence molecular orbitals (MOs) of consisting of delocalized $\pi$ electrons spread over the porphyrin ring's C and N atoms. These conjugated $\pi$ electronic systems are constructed by the linear combination of C-2p$_z$ and N-2p$_z$ atomic orbitals (see Figure \ref{fig_2}). The absorption spectrum of the free base H$_2$TPP 
mainly originates from singlet to singlet $\pi$-$\pi^*$ transitions within the porphyrin ring (see Figure \ref{fig_2}). Experimentally, two absorption bands are observed in the visible region, namely, Q$_x$ (1.91 - 2.07 eV or 600 - 650 nm) and Q$_y$ (2.25- 2.42 eV or 512 - 550 nm) bands.\cite{Gouterman1961SpectraPorphyrins, Edwards1971PorphyrinsPorphin, Ferrer1991ElectronicTetraphenylporphyrin, Manni2016CombiningMetal-Porphyrins} Experimentally, four peaks are observed for these Q$_x$ and Q$_y$ bands due to the symmetry breaking or vibrational coupling. The other band, generally referred to as the Soret band, always appears near the UV-region around 3.10 - 3.33 eV (400 - 372 nm), and presents the strongest absorption intensity.  

The NEVPT2 computed excitation energies using the (8,7) active space with 10 triplets and 10 singlet states (below 4 eV energy cut-off) efficiently capture all three excitation bands and display a nice agreement with the experimental values (see Table \ref{tab:Wavefunction decomposition of H2TPP}). Multiconfiguration wave function analysis reveals that the Q$_x$ and Q$_y$ bands originate from the electronic transition from the two doubly occupied $\pi$ orbitals to the two anti-bonding $\pi^*$ pair of orbitals (see Figure \ref{fig_2}). On the other hand, the high-energy Soret band arises from the dominant electronic transition from the lower-lying $\pi$-bonding orbitals (HOMO-2 or HOMO-3) to the first excited $\pi^*$ pairs. Interestingly, relative to the ground state singlet, the first excited state is found to be a triplet, located at 1.57 eV, compared to the first excited singlet, which is at 1.85 eV as obtained from the NEVPT2 calculations (see Table \ref{tab:Wavefunction decomposition of H2TPP}). The CASSCF electronic configurations (or occupations) shown in Table \ref{tab:Wavefunction decomposition of H2TPP} follow the same order as depicted in Fig \ref{fig_2}. Compared to NEVPT2 excitation energies, SA-CASSCF overestimates the excited states by more than 1 eV (see Tables S1 and S2 in SI). The (4,5) active space lacks the two lowest occupied MOs from the (8,7) one, thus missing the Soret band among its solutions. However, the Q$_x$ and Q$_y$ bands are predicted in a similar region as compared to (8,7) active space from NEVPT2 calculations (see Table S3). Notably, the Q$_x$ and Q$_y$ bands approach closer to each other when the acidic hydrogen atoms are removed to make it a dianion (see Table S4 in SI), in virtue of an overall higher symmetry of the complex.

\begin{table}[]
\centering
\caption{\textbf{Vertical excitation energies and wave function decomposition of H$_2$TPP from SA-CASSCF/NEVPT2 calculations. Only major determinants (u = spin up, d = spin down) are shown; conjugate determinants are omitted.}}
\label{tab:Wavefunction decomposition of H2TPP}
\renewcommand{\arraystretch}{1.1}
\resizebox{\columnwidth}{!}{%
\begin{tabular}{ccccc}
\hline
Singlet States & \begin{tabular}[c]{@{}c@{}}CASSCF Wave function\\ (major weightage)\end{tabular} & \begin{tabular}[c]{@{}c@{}}NEVPT2\\ (CASSCF) \\ energy (eV)\end{tabular} & \begin{tabular}[c]{@{}c@{}}Bands (NEVPT2\\ Osc. Str.)\end{tabular} & \begin{tabular}[c]{@{}c@{}}Exp. \\ (eV)\end{tabular} \\ \hline
GS & 2 2 2 2 0 0 0 (91\%) & 0.0 &  &  \\ \hline
1st ES & \begin{tabular}[c]{@{}c@{}}2 2 u 2 0 d 0 (49\%)\\ 2 2 2 u d 0 0 (43\%)\end{tabular} & 1.85 (2.88) & Q$_x$ (0.02) & \begin{tabular}[c]{@{}c@{}}1.91-\\ 2.07\end{tabular} \\ \hline
2nd ES & \begin{tabular}[c]{@{}c@{}}2 2 2 u 0 d 0 (49\%)\\ 2 2 u 2 d 0 0 (44\%)\end{tabular} & 2.29 (3.20) & Q$_y$ (0.01) & \begin{tabular}[c]{@{}c@{}}2.25-\\ 2.42\end{tabular} \\ \hline
3rd ES & \begin{tabular}[c]{@{}c@{}}u 2 2 2 d 0 0 (29\%)\\ 2 2 u 2 d 0 0 (28\%)\\ 2 2 2 u 0 d 0 (25\%)\end{tabular} & 2.98 (5.15) & Soret (1.46) & \multirow{4}{*}{\begin{tabular}[c]{@{}c@{}}3.10-\\ 3.33\end{tabular}} \\ \cline{1-4}
4th ES & \begin{tabular}[c]{@{}c@{}}u 2 2 2 0 d 0 (49\%)\\ 2 2 2 u d 0 0 (19\%)\end{tabular} & 3.00 (4.22) & Soret (0.2) &  \\ \cline{1-4}
5th ES & 2 u 2 2 0 d 0 (78\%) & 3.10 (4.19) & - &  \\ \cline{1-4}
6th ES & \begin{tabular}[c]{@{}c@{}}u 2 2 2 d 0 0 (46\%)\\ 2 2 u 2 d 0 0 (18\%)\end{tabular} & 3.17 (4.51) & Soret (0.2) &  \\ \hline
Triplet States &  &  &  &  \\ \hline
1st ES & \begin{tabular}[c]{@{}c@{}}2 2 u 2 u 0 0 (58\%)\\ 2 2 2 u 0 u 0 (35\%)\end{tabular} & 1.57 (2.42) &  &  \\ \hline
2nd ES & 2 2 u 2 0 u 0 (87\%) & 1.83 (2.85) &  &  \\ \hline
3rd ES & \begin{tabular}[c]{@{}c@{}}2 2 2 u 0 u 0 (60\%)\\ 2 2 u 2 u 0 0 (34\%)\end{tabular} & 1.91 (2.61) &  &  \\ \hline
4th ES & 2 2 2 u u 0 0 (90\%) & 2.15 (2.46) &  &  \\ \hline
\end{tabular}%
}
\renewcommand{\arraystretch}{1.0}
\end{table}

\textbf{Vanadyl tetra-phenyl porphyrin.} Next, we investigate the excited states of the VOTPP molecule. At first, the minimal active space ((1,5) or AS1) consists of only the 3d-metal orbitals used to compute the five excited states within the doublet spin multiplicity. The metal-centered orbitals are well separated due to the strong ligand field of the V(IV) ion, and the highest energy d$_{z{^2}}$ orbital lies 4.5 eV apart relative to the singly occupied d$_{xy}$ orbital (see Table S5 and Figure S1 in SI). Adding the second d-shell into the active space ((1,10) or AS2) barely changes the excitation energies of the VOTPP molecule (see Table S6 in SI). Expanding the active space onto the axial oxygen-based p$_x$, p$_y$, p$_z$ and the nitrogen-based $\sigma_{x^{2}-y^{2}}$ orbitals give rise to (9,9) active space which are also the bonding counterpart of the vanadyl d$_{xz}$, d$_{yz}$, d$_{z^{2}}$ and d$_{x^{2}-y^{2}}$ anti-bonding orbitals respectively (see Figure S2 in SI). The inclusion of these four bonding orbitals allows one to identify ligand-to-metal charge transfer (LMCT) states, which arise mainly from the overlap between oxygen-based $\pi_{xz}$/$\pi_{yz}$ and vanadium d$_{xz}$, d$_{yz}$, and d$_{xy}$ orbitals electronic transitions (see Table S7 in SI). These LMCT excited states appear from 5.4 eV and are significantly higher in energy than the d-d excited states. To mention, quartet LMCT excited states are lower in energy compared to the spin-allowed excited states.       


\begin{table*}[]
\centering
\caption{\textbf{Low-lying vertical excitation spectra and wavefunction decomposition of the VOTPP molecule from (13,14) active space calculations. Results are shown only up to 8 doublets and 4 quartets. Only major determinants (u = spin up, d = spin down) are shown; conjugate determinants are omitted.}}
\label{tab:Wavefunction_VOTPP}
\renewcommand{\arraystretch}{1.2}
\resizebox{\textwidth}{!}{%
\begin{tabular}{ccccc}
\hline
Doublet States &
  \begin{tabular}[c]{@{}c@{}}CASSCF wave function \\ (major weightage)\\ $\sigma_{z^2}$$\sigma_{{x^2}-{y^2}}$$\pi_{xz}$$\pi_{yz}$$\pi(p)$$\pi(p)$d$_{xy}$$\pi^*(p)$$\pi^*(p)$$\pi^*(p)$$\pi^*_{xz}$$\pi^*_{yz}$$\sigma^*_{{x^2}-{y^2}}$$\sigma^*_{z^2}$\end{tabular} &
  \begin{tabular}[c]{@{}c@{}}NEVPT2\\(CASSCF)\\ energy (eV)\end{tabular} &
  \begin{tabular}[c]{@{}c@{}}Bands \\ (Osc. Str.)\end{tabular} &
  \begin{tabular}[c]{@{}c@{}}Exp. \\ (eV)\end{tabular} \\ \hline
GS &
  {\spaceskip=11pt 2   2   2   2   2   2   u   0   0   0   0   0   0   0\ (83\%)} &
  0.0 &
   &
   \\ \hline
1st ES &
  {\spaceskip=11pt 2   2   2   2   u   2   u   0   d   0   0   0   0   0 (83\%)} &
  1.76 (2.94) &
  Q(10$^{-7}$) &
  \multirow{2}{*}{\begin{tabular}[c]{@{}c@{}}1.94-\\ 1.97\end{tabular}} \\ \cline{1-4}
2nd ES &
  {\spaceskip=11pt 2   2   2   2   u   2   u   d   0   0   0   0   0   0 (83\%)} &
  1.76 (2.94) &
  Q(10$^{-7}$) &
   \\ \hline
3rd ES &
  {\spaceskip=11pt 2   2   2   2   2   u   u   d   0   0   0   0   0   0 (84\%)} &
  2.07 (2.45) &
  Q(10$^{-7}$) &
  \multirow{2}{*}{\begin{tabular}[c]{@{}c@{}}2.14-\\ 2.18\end{tabular}} \\ \cline{1-4}
4th ES &
  {\spaceskip=11pt 2   2   2   2   2   u   u   0   d   0   0   0   0   0 (84\%)} &
  2.07 (2.45) &
  Q(10$^{-7}$) &
   \\ \hline
5th ES &
  \begin{tabular}[c]{@{}c@{}}{\spaceskip=11pt 2   2   2   2   2   d   u   u   0   0   0   0   0   0 (34\%)}\\ {\spaceskip=11pt 2   2   2   2   d   2   u   0   u   0   0   0   0   0 (25\%)}\end{tabular} &
  2.14 (3.14) &
  Q(0.02) &
  \multirow{2}{*}{\begin{tabular}[c]{@{}c@{}}2.25-\\ 2.30\end{tabular}} \\ \cline{1-4}
6th ES &
  \begin{tabular}[c]{@{}c@{}}{\spaceskip=11pt 2   2   2   2   2   d   u   0   u   0   0   0   0   0 (34\%)}\\ {\spaceskip=11pt 2   2   2   2   d   2   u   u   0   0   0   0   0   0 (25\%)}\end{tabular} &
  2.14 (3.14) &
  Q(0.02) &
   \\ \hline
7th ES &
  {\spaceskip=11pt 2   2   2   2   2   2   0   0   0   0   0   0   u   0 (83\%)} &
  2.65 (2.40) &
  - &
   \\ \hline
8th ES &
  {\spaceskip=11pt 2   2   2   2   2   u   0   u   0   0   0   0   d   0 (86\%)} &
  4.73 (4.86) &
  - &
   \\ \hline
9th ES &
  {\spaceskip=11pt 2   2   2   2   2   u   0   0   u   0   0   0   d   0 (86\%)} &
  4.73 (4.86) &
  - &
   \\ \hline
Quartet States &
   &
   &
   &
   \\ \hline
1st ES &
  {\spaceskip=11pt 2   2   2   2   u   2   u   0   u   0   0   0   0   0 (83\%)} &
  1.76 (2.93) &
   &
   \\ \hline
2nd ES &
  {\spaceskip=11pt 2   2   2   2   u   2   u   u   0   0   0   0   0   0 (83\%)} &
  1.76 (2.93) &
   &
   \\ \hline
3rd ES &
  {\spaceskip=11pt 2   2   2   2   2   u   u   u   0   0   0   0   0   0 (84\%)} &
  2.07 (2.45) &
   &
   \\ \hline
4th ES &
  {\spaceskip=11pt 2   2   2   2   2   u   u   0   u   0   0   0   0   0 (84\%)} &
  2.07 (2.45) &
   &
   \\ \hline
\end{tabular}%
}
\renewcommand{\arraystretch}{1.0}
\end{table*}

\begin{figure}[]
  \centering
  \vspace{5pt}
  \includegraphics[width=1.35\columnwidth, trim=130 10 10 10, clip]{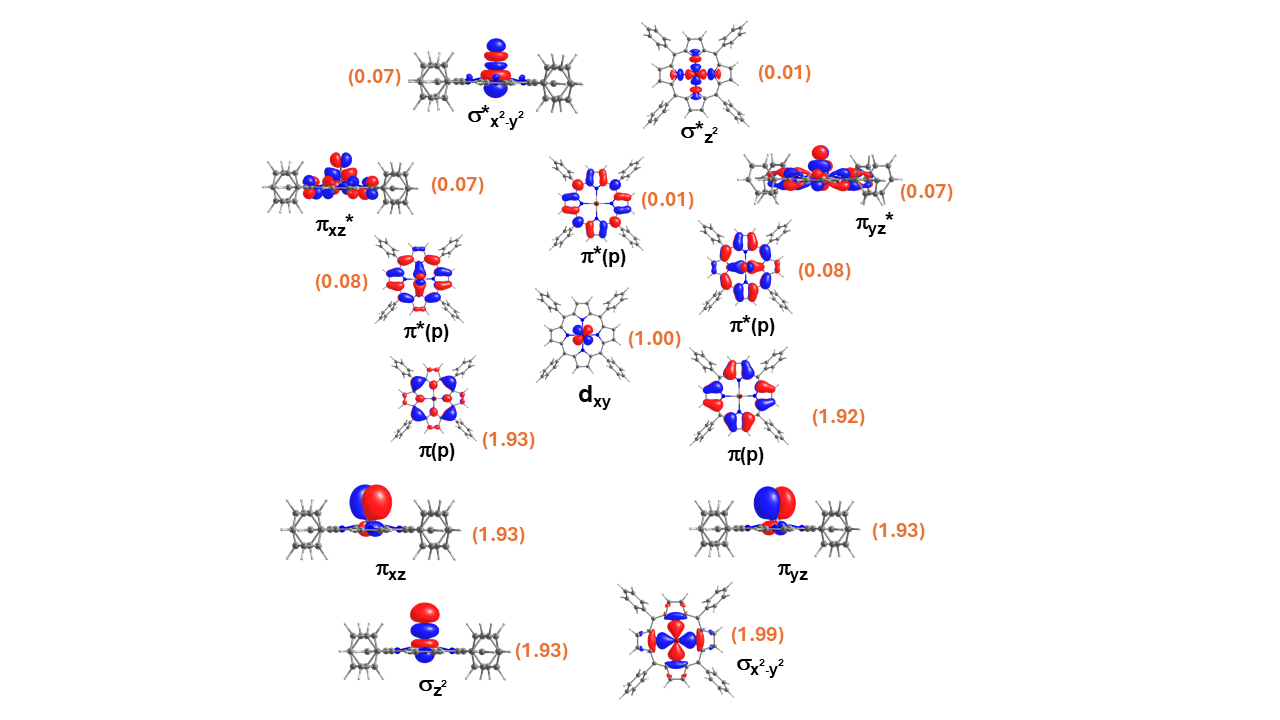}
  \caption{Natural orbitals of VOTPP obtained from state-specific CASSCF(13,14) calculations along with their natural orbital occupation numbers for the ground state.}
  \vspace{-10pt}
  \label{fig_3}
\end{figure} 

At this point, it is quite clear that the state-average calculations using (1,5) and (1,10) active spaces for VOTPP capture only the excited states based on the metal d-orbitals and do not address the low-lying excited states or Q-bands correspond to the $\pi$-$\pi^*$ excitations which are prominent in the H$_2$TPP molecule below 2 eV energy window. The (9,9) active space and the associated excited states, on the other hand, provide insights regarding the high-energy LMCT states, primarily localized on the V-O $\pi$-orbitals. However, none of these active spaces involve the porphyrin $\pi$ orbitals. Moreover, it is crucial to determine whether d-d excitations are the only spin-allowed and low-lying excited states that appear within 3 eV or if other ligand-centric ($\pi$-$\pi^*$ or MLCT) transitions also interfere with the d-d transition energies. In order to capture the complete ligand and metal-based electronic states, a large active space is constructed to address these questions by adding the delocalized ligand (or TPP) $\pi$ orbitals along with the existing V-O d/$\pi$-orbitals. Thus, the (13,14) active space is founded by combining the vanadyl (9,9) active space with the TPP (4,5) active space (see Figure \ref{fig_3}).  Both state-specific (with only ground state root) and state-average calculations have been performed for the VOTPP molecule (see Table \ref{tab:Wavefunction_VOTPP}). State-specific CASSCF(13,14) orbitals are shown in Figure \ref{fig_3} with the corresponding natural orbital occupations. It can be seen that the porphyrin $\pi$-$\pi^*$ orbitals are compressed between the V-N and V-O based $\sigma$ and $\pi$ bonding orbitals and the corresponding $\sigma^*$ and $\pi^*$ anti-bonding orbitals. Very importantly, the non-bonding d$_{xy}$ orbital, which carries the unpaired electron, is sandwiched between the five porphyrin ($\pi$/$\pi^*$) orbitals and remains localized on the V(IV) ion.

With these sets of active orbitals, state-average CASSCF calculations were carried out using nine doublet excited states and four quartet excited states to investigate the low-lying excitations. The state-average configurations (in determinant form) with their corresponding NEVPT2 vertical excitation energies are shown in Table \ref{tab:Wavefunction_VOTPP}. Notably, the quartet states in the VOTPP are now nearly degenerate with the low-lying doublet states- the first two quarters and doublets lie around 1.76 eV, and the third and fourth quartets and doublets lie at 2.07 eV. A closer inspection reveals that the energy splitting between the first two degenerate quartets and doublets are just 6-10 cm$^{-1}$, and the gap between the second degenerate quartets and doublets is 8 cm$^{-1}$, with the quartet pairs always being lower in energy. The significant distinction from this result of the (13,14) active space is that now the low-lying $\pi$-$\pi^*$ transitions are clearly visible, which were absent from the (1,5) and (9,9) active spaces. Indeed, the first six excited states from NEVPT2 are the porphyrin $\pi$-$\pi^*$ transitions, thereby giving rise to the Q$_x$ and Q$_y$ bands. It is apparent from the comparison of H$_2$TPP spectra (see Tables \ref{tab:Wavefunction decomposition of H2TPP} and \ref{tab:Wavefunction_VOTPP}) that the incorporation of the vanadyl cation results in a red shift (lowering in energy) of the Q$_x$ and Q$_y$ bands, in agreement with the experimental observations\cite{Ferrer1991ElectronicTetraphenylporphyrin}. The SA-CASSCF(13,14) results, on the other hand, unsurprisingly overestimate the excited state energies (see Table S8 in SI). Also, there is one metal-based d$_{xy}$-d$_{x^{2}-y^{2}}$ transition, which is expected to lie higher in energy compared to the $\pi$-$\pi^*$ transitions is incorrectly identified from the CASSCF calculations (see Tables \ref{tab:Wavefunction_VOTPP} and S8). Previous photophysical studies on VOTPP and related molecules \cite{doi:10.1021/jacs.4c10632} support the presence of low-lying quartet states \cite{Harima1997PhosphorescenceTemperature}.

\textbf{Copper tetra-phenyl porphyrin.}
Next, we perform the active space exploration for the CuTPP molecule. As usual, the metal-based (9,5) is examined first. The d-orbitals occupation and the corresponding SA-CASSCF and NEVPT2 excitation energies are reported in Table S9 and Figure S3 in SI. Despite being a one-electron (or one hole) system, there is a clear difference between the V(IV) and Cu(II) d-orbital splitting within the TPP ligand environment. Ab initio ligand field analysis (AI-LFT from NEVPT2 energies) revealed that the ligand-field splitting (difference between the lowest and highest d-orbital energies) in the Cu(II) center is around 2.2 eV, which is more than 2 eV smaller compared to the V(IV) center (see Tables S5 and S8 in SI). This is expected as vanadium is in a $+4$ oxidation state and additionally bonded to a very electronegative oxygen atom, while copper is in a $+2$ oxidation state and not bonded to any electronegative species in the axial direction. Importantly, when the second d-shell is incorporated into the active space for CuTPP, the vertical excitation energies of the first five doublets from NEVPT2(9,10) are reduced by $\sim$0.4 eV and SA-CASSCF excited states are increased by $\sim$0.1 eV (see Tables S9 and S10 in SI). Interestingly, this is contrary to what we observe for VOTPP,  where double-shell minutely affects the energies of the five doublet energies (see Tables S5 and S6 in SI).

\begin{figure}[]
  \centering
  \vspace{5pt}
  \includegraphics[width=1.30\columnwidth, trim=150 10 10 10, clip]{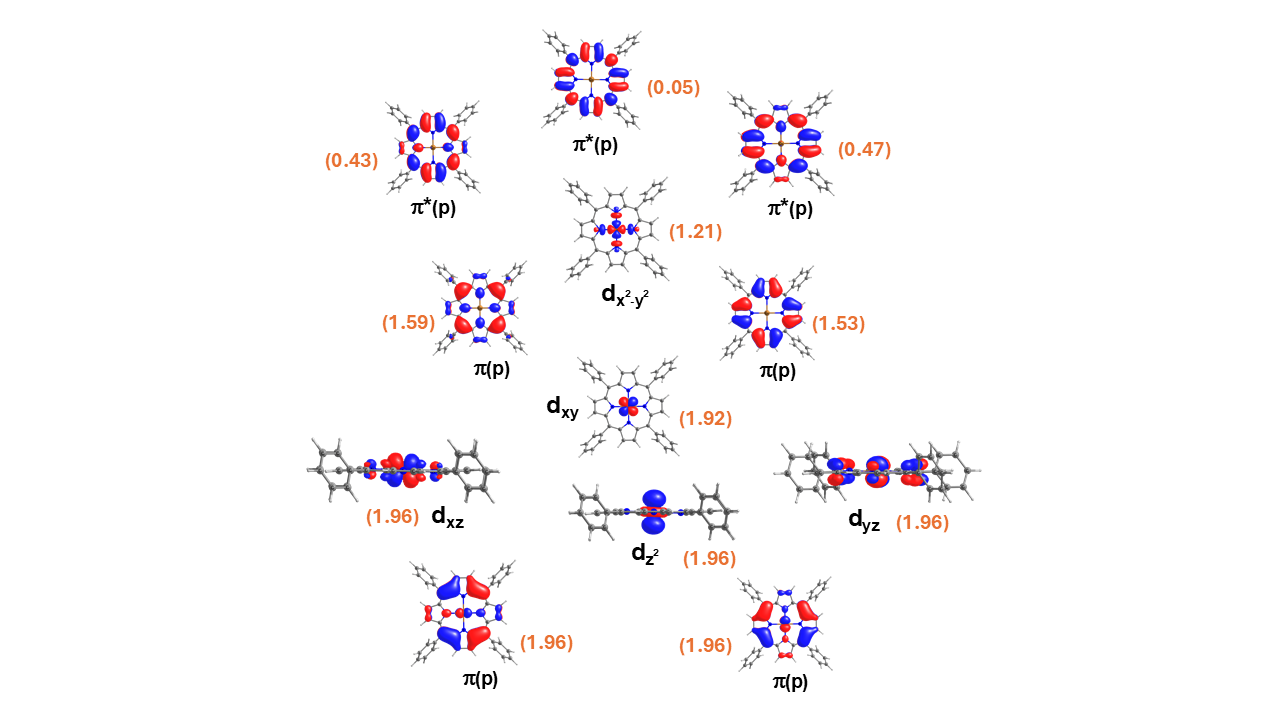}
  \caption{Active orbitals of CuTPP obtained from SA-CASSCF(17,12) calculations using 12 doublets and 4 quartets. The numbers in orange correspond to their pseudo-natural orbital occupation numbers.}
  \vspace{-10pt}
  \label{fig_4}
\end{figure} 

\begin{table*}[ht]
\centering
\caption{\textbf{Wavefunction decomposition and vertical excited states of the CuTPP molecule from (17,12) active space calculations. Only major determinants (u = spin up, d = spin down) are shown, conjugate determinants are omitted}.}
\label{tab:Wavefunction_CuTPP}
\renewcommand{\arraystretch}{1.2}
\resizebox{\textwidth}{!}{%
\begin{tabular}{ccccc}
\hline
Doublet States & \begin{tabular}[c]{@{}c@{}}CASSCF wave function \\ (major weightage)\\ $\pi(p)$$\pi(p)$d$_{z^2}$d$_{xz}$d$_{yz}$d$_{xy}$ $\pi(p)$$\pi(p)$d$_{{x^2}-{y^2}}$$\pi^*(p)$$\pi^*(p)$$\pi^*(p)$\end{tabular} & \begin{tabular}[c]{@{}c@{}}NEVPT2\\(CASSCF)\\energy(eV)\end{tabular} & \begin{tabular}[c]{@{}c@{}}Bands \\ (Osc. Str.)\end{tabular} & \begin{tabular}[c]{@{}c@{}}Exp.\\ (eV)\end{tabular} \\ \hline
GS & {\spaceskip=12pt 2 2 2 2 2 2 2 2 u 0 0 0 (88\%)} & 0.0 &  &  \\ \hline
1st ES & {\spaceskip=12pt 2 2 2 2 2 2 u 2 u 0 d 0 (88\%)} & 1.84 (2.70) &  &  \\ \hline
2nd ES & {\spaceskip=12pt 2 2 2 2 2 2 u 2 u d 0 0 (88\%)} & 1.85 (2.68) &  &  \\ \hline
3rd ES & {\spaceskip=12pt 2 2 2 2 2 u 2 2 2 0 0 0 (88\%)} & 2.03 (1.46) &  &  \\ \hline
4th ES & {\spaceskip=12pt 2 2 2 2 2 2 2 u u 0 d 0 (89\%)} & 2.12 (2.35) & Q(10$^{-6}$) & \multirow{2}{*}{2.00} \\ \cline{1-4}
5th ES & {\spaceskip=12pt 2 2 2 2 2 2 2 u u d 0 0 (89\%)} & 2.13 (2.32) & Q(10$^{-6}$) &  \\ \hline
6th ES & \begin{tabular}[c]{@{}c@{}}{\spaceskip=12pt 2 2 2 2 2 2 2 u u d 0 0 (51\%)}\\ {\spaceskip=12pt 2 2 2 2 2 2 u 2 u 0 d 0 (36\%)}\end{tabular} & 2.16 (2.98) & Q(0.02) & \multirow{2}{*}{2.30} \\ \cline{1-4}
7th ES & \begin{tabular}[c]{@{}c@{}}{\spaceskip=12pt 2 2 2 2 2 2 2 u u 0 d 0 (51\%)}\\ {\spaceskip=12pt 2 2 2 2 2 2 u 2 u d 0 0 (37\%)}\end{tabular} & 2.16 (2.96) & Q(0.02) &  \\ \hline
8th ES & \begin{tabular}[c]{@{}c@{}}{\spaceskip=12pt 2 2 2 2 u 2 2 2 2 0 0 0 (67\%)}\\ {\spaceskip=12pt 2 u 2 2 2 2 2 2 2 2 0 0 (21\%)}\end{tabular} & 2.16 (1.72) &  &  \\ \hline
9th ES & \begin{tabular}[c]{@{}c@{}}{\spaceskip=12pt 2 2 2 u 2 2 2 2 2 0 0 0 (67\%)}\\ {\spaceskip=12pt u 2 2 2 2 2 2 2 2 2 0 0 (09\%)}\end{tabular} & 2.17 (1.72) &  &  \\ \hline
10th ES & {\spaceskip=12pt 2 2 u 2 2 2 2 2 2 0 0 0 (88\%)} & 2.27 (1.81) &  &  \\ \hline
Quartet States &  &  &  &  \\ \hline
1st ES & {\spaceskip=12pt 2 2 2 2 2 2 u 2 u u 0 0 (88\%)} & 1.82 (2.67) &  &  \\ \hline
2nd ES & {\spaceskip=12pt 2 2 2 2 2 2 u 2 u 0 u 0 (88\%)} & 1.83 (2.67) &  &  \\ \hline
3rd ES & {\spaceskip=12pt 2 2 2 2 2 2 2 u u 0 u 0 (89\%)} & 2.11 (2.34) &  &  \\ \hline
4th ES & {\spaceskip=12pt 2 2 2 2 2 2 2 u u u 0 0 (89\%)} & 2.12 (2.31) &  &  \\ \hline
\end{tabular}%
}
\renewcommand{\arraystretch}{1.0}
\end{table*}

An intermediate active space of (11,6) can be constructed for CuTPP by incorporating the doubly occupied $\sigma_{{x^2}-{y^2}}$ orbital into the (1,5) active space consisting of the nitrogen coordinating orbitals. State average calculations using five doublet roots using this active space in fact, reinstates the (9,5) active space results both from SA-CASSCF and NEVPT2 energies (see Table S11 and Figure S4 in SI). This can be attributed to the more balanced electron correlation by including the Cu-N bonding ($\sigma_{x{^2}-y{^2}}$)and the anti-bonding $\sigma^*_{{x^2}-{y^2}}$ orbitals compared to the (9,10) active space where the Cu 3d-electrons gained additional delocalization.

Analogous to the VOTPP, a large active space of (17,12) was constructed for the CuTPP molecule. This has been achieved by combining the (4,5) active space of the TPP moiety, (9,5) of the Cu(II) center and a degenerate pair of doubly occupied porphyrin $\pi$ orbitals. We detected that the nitrogen p$_z$ orbitals contained in this degenerate pair of $\pi$ orbitals weakly interact with the d$_{xz}$/d$_{yz}$ orbitals of the Cu(II) center in the CuTPP molecule. Contrary to the VOTPP case, this degenerate $\pi$ pair of orbitals was unavailable to accommodate for the VOTPP case as the d$_{xz}$/d$_{yz}$ orbitals of the V(IV) center were involved in strong bonding interaction with the oxygen atom. Attempts to optimize the active orbitals only for the ground state were unsuccessful; therefore, at first, state-average calculations were performed using five roots. This may be attributed to the close-lying five excited states originating from the five d-d excitations, which were found to lie within 14000 cm$^{-1}$ obtained from SA-CASSCF energies (see Table S12 in SI). In fact, the d-orbitals in the CuTPP from the SA(5)-CASSCF calculation all possess a natural orbital occupation of 1.8, indicating a strong electron correlation within the five orbitals (see Figure S5 in SI). However, this lacks complete electron correlation and low-lying $\pi$-$\pi^*$ transitions, and hence, the SA(5)-CASSCF orbitals were further optimized using twelve doublets and four quartets. The state-average orbitals are plotted in Figure \ref{fig_4} with their state-average pseudo-natural occupations, and the excited state configurations with their NEVPT2 energies shown in Table \ref{tab:Wavefunction_CuTPP}. Finally, by using a fair amount of state average roots, NEVPT2 results ascertain that the first four excited states originate from the porphyrin $\pi$-$\pi^*$ transitions (see Table \ref{tab:Wavefunction_CuTPP}). Similar to VOTPP, in the case of CuTPP, the first two doublet pairs and the first two quartet pairs are quite close in energy, and the energy difference between the first quartet pair and the first doublet pair is roughly 170 cm$^{-1}$, with quartet pairs being lower in energy. The next two doublet pairs at 2.12-2.13 eV (585 nm) and 2.16 eV show significant oscillator strength from NEVPT2 calculations, and are in close agreement with the experimental absorption spectra,\cite{Ralphs2017SolventlessPorphyrins, Al-Shewiki2019Diaqua--octaferrocenyltetraphenylporphyrin:Species, Chernyadyev2022CopperIICompounds}, which is detected around 2.00 eV (620 nm) and 2.30 eV (540 nm) respectively. The SA-CASSCF results, on the other hand, predict the d-d excited states to be lower in energy compared to the porphyrin $\pi$-$\pi^*$ transitions (see Table S13 in SI). The $\pi$-$\pi^*$ transitions are also overestimated from the SA-CASSCF calculations, which further emphasizes the role of second-order perturbative corrections in determining the excited state spectra of these types of molecules.

\textbf{Spin-orbit coupling and $\mathbf{g}$-shifts.} For all the active spaces discussed above for VOTPP and CuTPP, \textit{g}-tensors are computed using QDPT (SOC) and an effective Hamiltonian approach (EHA) for the ground state Kramers pair (see Table \ref{tab:g-factor}). The largest contribution to the $g$-value in the case of transition metal complexes is from the spin-orbit/Zeeman interaction, which is represented within a perturbation theory approach by\cite{Singh2018ChallengesComplexes}
\begin{equation}
\scalebox{0.71}{$
\begin{aligned}
    g_{KL}^{(\text{OZ/SOC})} = & - \frac{1}{S} \sum_{b(S_b = S)} \Delta_b^{-1}
    \Biggl\{\left\langle \Psi_0^{SS} \left| \sum_i l_{i;K} \right| \Psi_b^{SS} \right\rangle
    \left\langle \Psi_b^{SS} \left| \sum_i z_{L;i} s_{z;i} \right| \Psi_0^{SS} \right\rangle \\
    & + \left\langle \Psi_0^{SS} \left| \sum_i z_{K;i} s_{z;i} \right| \Psi_b^{SS} \right\rangle
    \left\langle \Psi_b^{SS} \left| \sum_i l_{i;L} \right| \Psi_0^{SS} \right\rangle\Biggr\} \:.
\end{aligned}
$}
\label{eq:1}
\end{equation}

Here, $S$ represents the ground state spin of the species, 1/2 in out case, $\Psi^{SS}_0$ and $\Psi^{SS}_b$ represent the ground state wave function and the excited state wave function within the same $S$ multiplet. $l_{i;K}$ and $l_{i;L}$ are the \textit{K}-th and \textit{L}-th components of the orbital angular momentum of the electron. Here, $z_{L;i}$ is the L-th component of the effective one-electron obtained from the SOMF approximation to the full SOC operator.
For the vanadyl, $g_z$ is expected to be small compared to $g_x$ and $g_y$ as the numerator of equation \ref{eq:1} is larger for the transition d$_{xy}$-d$_{{x^2}-{y^2}}$ compared to d$_{xy}$-d$_{xz}$/d$_{yz}$ transitions. Except for the large (13,14) active space of VOTPP, the other active spaces include all five d-d excitations, which are essentially required to perturb the three \textit{g}-factors from the free electron value of 2.0023. For the (13,14) active space, ten doublet roots are not enough to incorporate all the d-d excitations except the d$_{xy}$-d$_{{x^2}-{y^2}}$ transition and hence, both the $g_x$ and $g_y$ values for this active space are close to 2.0023. Among the results of NEVPT2 (1,5) and NEVPT2 (1,10), no significant differences are observed in the \textit{g} factors. Inclusion of the LMCT states, that is, in the (9,9) active space, significantly improves the $g$-factors in comparison to experimental values, particularly for $g_x$ and $g_y$. 

\begin{table}[t]
\centering
\caption{Comparison of the calculated and experimental g-factors of VOTPP and CuTPP molecules. Experimental values were taken from ref. \cite{Yamabayashi2018ScalingFramework} for VOTPP and ref.\cite{Urtizberea2018ANanosheets} for CuTPP.}
\label{tab:g-factor}
\resizebox{\columnwidth}{!}{%
\begin{tabular}{ccccc}
\hline
Species & \begin{tabular}[c]{@{}c@{}}Method/\\ Number of excited states\end{tabular}            & g$_x$    & g$_y$    & g$_z$    \\ \hline
\multirow{10}{*}{VOTPP} & \begin{tabular}[c]{@{}c@{}}SA-CASSCF(1,5)\\ 5 doublets\end{tabular} & 1.984 & 1.984 & 1.934 \\ \cline{2-5} 
        & \begin{tabular}[c]{@{}c@{}}NEVPT2(1,5)\\ 5 doublets\end{tabular}                      & 1.988 & 1.988 & 1.947 \\ \cline{2-5} 
        & \begin{tabular}[c]{@{}c@{}}SA-CASSCF(1,10)\\ 5 doublets\end{tabular}                  & 1.984 & 1.984 & 1.934 \\ \cline{2-5} 
        & \begin{tabular}[c]{@{}c@{}}NEVPT2(1,10)\\ 5 doublets\end{tabular}                     & 1.988 & 1.988 & 1.946 \\ \cline{2-5} 
        & \begin{tabular}[c]{@{}c@{}}SA-CASSCF(9,9)\\ 5 doublets\end{tabular}                   & 1.989 & 1.989 & 1.939 \\ \cline{2-5} 
        & \begin{tabular}[c]{@{}c@{}}NEVPT2(9,9)\\ 5 doublets\end{tabular}                      & 1.986 & 1.986 & 1.946 \\ \cline{2-5} 
        & \begin{tabular}[c]{@{}c@{}}SA-CASSCF(9,9)\\ 9 doublets\end{tabular}                   & 1.988 & 1.988 & 1.940 \\ \cline{2-5} 
        & \begin{tabular}[c]{@{}c@{}}NEVPT2(9,9)\\ 9 doublets\end{tabular}                      & 1.987 & 1.987 & 1.948 \\ \cline{2-5} 
        & \begin{tabular}[c]{@{}c@{}}SA-CASSCF(13,14)\\ 10 doublets and 4 quartets\end{tabular} & 2.002 & 2.002 & 1.938 \\ \cline{2-5} 
        & \begin{tabular}[c]{@{}c@{}}NEVPT2 (13,14)\\ 10 doublets and 4 quartets\end{tabular}   & 2.002 & 2.002 & 1.944 \\ \hline
Exp     &                                                                                       & 1.986 & 1.986 & 1.963 \\ \hline
\multirow{8}{*}{CuTPP}  & \begin{tabular}[c]{@{}c@{}}SA-CASSCF(9,5)\\ 5 doublets\end{tabular} & 2.097 & 2.098 & 2.501 \\ \cline{2-5} 
        & \begin{tabular}[c]{@{}c@{}}NEVPT2(9,5)\\ 5 doublets\end{tabular}                      & 2.078 & 2.078 & 2.347 \\ \cline{2-5} 
        & \begin{tabular}[c]{@{}c@{}}SA-CASSCF(9,10)\\ 5 doublets\end{tabular}                  & 2.097 & 2.097 & 2.491 \\ \cline{2-5} 
        & \begin{tabular}[c]{@{}c@{}}NEVPT2(9,10)\\ 5 doublets\end{tabular}                     & 2.093 & 2.094 & 2.446 \\ \cline{2-5} 
        & \begin{tabular}[c]{@{}c@{}}SA-CASSCF(11,6)\\ 5 doublets\end{tabular}                  & 2.098 & 2.098 & 2.502 \\ \cline{2-5} 
        & \begin{tabular}[c]{@{}c@{}}NEVPT2(11,6)\\ 5 doublets\end{tabular}                     & 2.079 & 2.079 & 2.351 \\ \cline{2-5} 
        & \begin{tabular}[c]{@{}c@{}}SA-CASSCF(17,12)\\ 5 doublets\end{tabular} & 2.097 & 2.098 & 2.500 \\ \cline{2-5} 
        & \begin{tabular}[c]{@{}c@{}}NEVPT2(17,12)\\ 5 doublets\end{tabular} & 2.078 & 2.078 & 2.347 \\ \cline{2-5} 
        & \begin{tabular}[c]{@{}c@{}}SA-CASSCF(17,12)\\ 12 doublets and 4 quartets\end{tabular} & 2.089 & 2.089 & 2.478 \\ \cline{2-5} 
        & \begin{tabular}[c]{@{}c@{}}NEVPT2(17,12)\\ 12 doublets and 4 quartets\end{tabular}    & 2.076 & 2.076 & 2.349 \\ \hline
Exp     &                                                                                       & 2.065 & 2.065 & 2.200 \\ \hline
\end{tabular}%
}
\end{table}

Due to the presence of pseudo-D$_{4h}$ symmetry, the $g_x$ and $g_y$ components are similar for both molecules, and are usually referred to as $g_\perp$. In the case CuTPP, $g_z$ is larger than the $g_x$ and $g_y$ values as the relative splitting between the d$_{xy}$-d$_{x{^2}-y{^2}}$ orbitals is smaller in the case of CuTPP than from VOTPP. Here, a clear difference is visible between the NEVPT2(9,5) and NEVPT2(9,10) $g$-factors, since the spin-orbit free excitation energies were underestimated in the (9,10) active space from NEVPT2(9,10) method, which significantly overestimates the computed $g$-factors. This observation has previously been encountered in the estimation of $g$-factors using NEVPT2(9,10) methods.\cite{Singh2018ChallengesComplexes} The (11,6) active space, on the other hand, outperforms the (9,10) active space in this regard, providing reasonable $g$-factors for the CuTPP molecule. Unlike VOTPP, in the case of CuTPP, the largest active space (17,12) with twelve doublets includes all four d-d excited states and hence exhibits satisfactory $g$-factors compared to the other active space results. In fact, the SA-CASSCF(17,12) and NEVPT2(17,12) computed $g$-factors are comparable to those (9,5) active space ones.   

\section*{Discussion and conclusions} 

Four different active spaces have been chosen to inspect the ligand-to-metal charge transfer (LMCT), ligand-centric $\pi$-$\pi^*$ and metal-centric d-d excitations in VOTPP and CuTPP molecular qubits. For the free H$_2$TPP molecule, (8,7) active space captures all the required Q$_x$/Q$_y$ and Soret bands in order to explain the experimental absorption spectra compared to the smaller (4,5) active space, which is insufficient to determine the high-intensity Soret bands. For VOTPP and CuTPP, the minimal ((1,5) or (9,5)) and the double-shell embedded ((1,10), or (9,10)) active spaces primarily account for the d-d excitations but do not provide a full description of the complete spin state energies and do not align with the experimental absorption spectra. However, these active spaces yield a very accurate estimation of $g$-factors. Intermediate active spaces like (9,9) for VOTPP or (11,6) for CuTPP, which account for the metal-ligand bonding interaction up to the first coordination shell, improve the $g$-values to some extent but are insufficient to reproduce the low-lying $\pi$-$\pi^*$ transitions. Largest active space models like (13,14) and (17,12), on the other hand, quantitatively describe the low-lying $\pi$-$\pi^*$ excited states efficiently for both molecules. Largest active space calculations confirm that both for the VOTPP and CuTPP molecules, the low-lying excited states are dominated by quartet states. Our study thus makes an important step in correctly representing the low-lying excited states of this central class of molecular qubits, and paves the way to an investigation of the role of these overlooked electronic features in the spin-lattice relaxations and photophysical properties of spin-1/2 molecules. 
\vspace{0.2cm}

\noindent
\textbf{Acknowledgements and Funding}\\
A.L. acknowledges funding from the European Research Council (ERC) under the European Union’s Horizon 2020 research and innovation programme (grant agreement No. [948493]), and A.S. from the Marie Sklodowska Curie action (grant agreement No. 101151501). Trinity College Research IT and the Irish Centre for High-End Computing (ICHEC) provided computational resources.

\vspace{0.2cm}
\noindent
\textbf{Data Availability}\\
All data needed to evaluate the conclusions of the work are present in the paper and the Supplementary Information.

\vspace{0.2cm}
\noindent
\textbf{Conflict of interests}\\
The authors declare no competing interests.

\bibliographystyle{achemso}
\bibliography{Bibliography/references}

\end{document}